\documentclass[useAMS,usenatbib,usegraphicx,epsfig]{mn2e}
\usepackage{amssymb}
\usepackage{graphicx}
\usepackage{amsmath}

\newcommand{\rmin}{$r_{min}$}
\newcommand{\rmax}{$r_{max}$}
\newcommand{\lcdm}{\ensuremath{\Lambda\mathrm{CDM}}}

\title
[Redshift Recovery]
{Recovering Redshift Distributions with Cross-Correlations: Pushing The Boundaries}

\author[Schmidt et al.]
{Samuel J.~Schmidt$^{1}$\thanks{Email: sschmidt@physics.ucdavis.edu},
Brice M\'{e}nard$^{2}$\thanks{Alfred P.~Sloan fellow},
Ryan Scranton$^{1}$, 
Christopher Morrison$^{1}$, 
\newauthor Cameron K. McBride$^{3}$\\
$^{1}$Department of Physics, University of California, One Shields Avenue, Davis, CA 95616, USA \\
$^{2}$Department of Physics and Astronomy, Johns Hopkins University, 3400 N. Charles Street, Baltimore, MD, USA\\
$^{3}$Harvard–-Smithsonian Center for Astrophysics, 60 Garden Street, Cambridge, MA 02138, USA}

\begin{document}

\maketitle

\begin{abstract}
Determining accurate redshift distributions for very large samples of objects has become increasingly important in cosmology.  We investigate the impact of extending cross-correlation based redshift distribution recovery methods to include small scale clustering information.  The major concern in such work is the ability to disentangle the amplitude of the underlying redshift distribution from the influence of evolving galaxy bias.  Using multiple simulations covering a variety of galaxy bias evolution scenarios, we demonstrate reliable redshift recoveries using linear clustering assumptions well into the non-linear regime for redshift distributions of narrow redshift width.
Including information from intermediate physical scales balances the increased information available from clustering and the residual bias incurred from relaxing of linear constraints.  We discuss how breaking a broad sample into tomographic bins can improve estimates of the redshift distribution, and present a simple bias removal technique using clustering information from the spectroscopic sample alone.  
\end{abstract}


\section{Introduction}\label{intro}
Cosmological measurements require distance estimates in order to map the large scale structure of the universe.  In the past this has most often been done on an object by object basis by obtaining spectroscopic redshifts for individual sources.  Surveys of large samples of galaxies such as the Sloan Digital Sky Survey and its extensions \citep[]{York:00,sdss} have been instrumental in improving cosmological measurements.  However, a number of current and upcoming missions (DES, LSST, etc...) will attempt to measure the fundamental properties of cosmology, and particularly dark energy, using a variety of methods (e.g. weak gravitational lensing, baryon acoustic oscillations, etc.).  Fundamental to all of these surveys is the assumption that the millions, or even billions,  
of galaxies observed by these instruments will be separable into redshift bins, despite the fact that the number of objects involved makes spectroscopic follow-up wildly impractical.  Photometric redshift techniques show a good deal of promise towards this goal \citep[e.~g.~][]{Con:95,Ben:00,Cun:09}, but there remain questions
as to whether or not they can meet the stringent requirements outlined in these surveys and avoid systematic biases that could leak into dark energy constraints \citep[e.~g.~][]{Ma:06,Cun:12}.  
In this paper, we examine a  technique that uses clustering between spectroscopic and photometric samples to accurately determine a photometric sample's redshift distribution.  The applications of such a technique are much more general than the aforementioned large surveys: this technique can be used to estimate the redshift distribution of nearly any data set.  Even single-band detections that lack photometric redshift estimates can be used, as long as they have reliable astrometric information for the calculation of cross-correlation functions.  

The technique described in this paper uses the physical associations due to large scale clustering to probe redshift distributions.  Such ideas are not new:
\citet[]{Seld:79} cross-correlated quasars and galaxy counts to test for physical association, though they found no trend with redshift.  \citet[]{Rob:79} similarly cross-correlated quasars and rich galaxy clusters.  More recently, \citet[]{Qua:10} counted pairs of galaxies at small angular separations between photometric redshift selected samples, taking advantage of physically associated pairs of galaxies in order to determine an empirical measure of the photometric redshift errors.  Similarly, \citet[]{Ben:10} cross-correlated photometric redshift bins to determine the relative contamination fraction between redshift bins based on the magnitude of the induced angular cross-correlations.  

These previous techniques do not require any spectroscopic sample and rely solely on photometric redshift information.  \citet[]{Sch:06} discuss using cross-correlations of objects sorted into redshift bins in order to determine their redshift distribution.  They mention that having a subset of objects with more accurately determined redshifts would enable tighter constraints than photometric redshifts alone.  Expanding on this idea, \citet[]{New:08} (hereafter N08) and \citet[]{Matt:10,Matt:11} describe a technique that requires a spectroscopic sample that spans the redshift range of interest.  In simple terms, the method measures the amount of overlap between the spectroscopic sample divided into redshift bins and an unknown sample (which we will refer to as the ``photometric'' sample, though photometric redshifts are not necessary for sample selection).  As galaxies cluster on all scales, if a spectroscopic bin overlaps in redshift with the photometric sample, we expect to see an excess number of objects, whereas if there is no overlap we expect to simply see the average number of objects that overlap spatially due to projection.  By measuring the strength of the spatial cross-correlations as a function of redshift we can recover the redshift distribution of the photometric sample.
A major component of the N08 technique is an iterative method to correct for bias evolution that may occur in the sample.  \citet[]{Sch:10} implemented a very similar technique on mock data and reported difficulty in distentangling the galaxy bias from the redshift distribution, a point which we will examine in this paper.  It is only very recently that this technique has been used with real data \citep[]{Mit:12,Nik:12}, including the exact technique described in this work \citep[][ M\'{e}nard et al, in preparation]{Mor:12}.

The N08 technique is designed to work with large scale correlations where the galaxy bias can be treated as linear.  However, the increasing amount of  power in galaxy correlation functions due to large scale clustering  means that there is considerable signal that is not being fully utilized at  smaller scales (this is particularly relevant given the small angular  extent of many deep spectroscopic surveys).  Further study of the  spectroscopic sample's non-linear bias properties may enable us to  account for the effects of bias in the redshift recovery.  In this paper we explore the impact of retaining the linear assumptions while expanding the procedure to include smaller physical scales, where the galaxy bias becomes non-linear in the density field.  In this regime evolution of the galaxy bias will modulate the amplitude of the  recovered redshift distribution.  We test the efficacy of using the linear assumptions well into the non-linear regime.   

Additionally,  in some instances we are only interested in the existence or absence of galaxies in a redshift interval, and the detailed shape of the redshift distribution is  not the main concern. As an example, when selecting objects based on photometric redshifts, parameter degeneracies can lead to inclusion of a secondary population of objects far outside the intended redshift range.  In such a case the exact shape of the redshift distribution, which can be distorted by the presence of evolving galaxy bias, may be secondary to detecting the presence or absence of an interloper population.  For these reasons we examine the relative amount of  information contained at a range of physical scales around our  spectroscopic samples.

The layout of the paper is as follows: in \S\,\ref{method} we discuss the algorithms used to determine the redshift distributions.  A summary of the mock datasets is given in \S\,\ref{sims}.  Results are presented in \S\,\ref{results}.  We conclude and present future work in \S\,\ref{future}.

\section{Method}\label{method}

Our main goal is to measure the redshift probability density function (pdf), $\phi(z)$, for a specific sample of objects that we  will refer to as the ``photometric'' sample.  
As mentioned in Section~\ref{intro}, we do not necessarily need a photometric redshift measurement for our samples.  However, as we will discuss, a sample that is selected to cover a narrow range in redshift leads to a more accurate recovery than that of a broad distribution.  So, while the method can be applied to almost any arbitrary data set, in practice it will most often use samples selected with photometric redshifts.
We estimate these redshift distributions by measuring the amplitude of the cross-correlation signal between our photometric sample and a sample of objects with known redshifts.

The angular cross-correlation between the photometric sample and spectroscopic sample, $w_{sp}(\theta,z)$ is defined in terms of the mean density of objects an anglular distance $\theta$ from objects in the spectroscopic sample:
\begin{equation}\label{sigmaeqn}
\langle \Sigma(\theta,z) \rangle = \Sigma_{p}(1 + w_{sp}(\theta,z))
\end{equation}
\noindent where $\Sigma_{p}$ is the mean surface density of photometric objects.
In practice, rather than measuring the correlation function in multiple angular bins and fitting a power law form, we measure the density of ``photometric'' sources in a single physical annulus around each individual spectroscopic source, from a minimum  radius (\rmin) to a maximum radius (\rmax), measured in units of comoving $kpc$.   We subdivide the spectroscopic sample into bins of redshift and measure the mean (over)density of ``photometric'' objects around each spectroscopic source  within each redshift bin.   After subtraction of the average density expected from points randomly  placed in the survey geometry and normalization, this is equivalent to  the Natural Estimator, $DD/RR\,-\,1$ \citep[]{Ker:00}, where $DD$ is the number of cross-correlated pairs within our annulus and $RR$ is the number of correlated pairs from a dataset with randomized positions in the same survey footprint.  We use the amplitude of this ``one bin'' estimate of the excess clustering as our estimator of $\phi(z)$.  In addition to calculating the density with uniform weight within the annulus, we also also compute a density measure where we weight each object proportional to the inverse of the spatial distance from the spectroscopic object.  We will compare these estimators in the Appendix.  All calculations in  the body of the paper will use the inverse weighted estimator.

In computing the projected overdensities, proper treatment of the survey  area, including complicated selection and masks, is essential.  To accomplish this goal we develop code which  employs the {\it astro-stomp} software package\footnote{available at:  http://code.google.com/p/astro-stomp/}.  The software uses a  pixelization of the sky to encode both the galaxy positions as well as  the survey footprint for fast computation of galaxy density and has the  ability to encode complex masking and selection. 

\subsection{Bias Correction}\label{bias}
Measuring the overdensity of galaxies around objects in the spectroscopic sample does not  immediately give us the underlying redshift distribution: cross-correlations measure the object overdensity within a fixed real space annulus, but the  clustering length of both the spectroscopic and ``photometric'' samples are  not necessarily constant with redshift.  We must account for any such evolution in order to recover the redshift distribution.
We will examine using the clustering length calculated from the spectroscopic sample alone to account for bias evolution in Section~\ref{approx}, but will mainly use the iterative method introduced by N08.
This technique describes an iterative correction to the redshift distribution using estimates of the mean clustering length for the photometric sample and the (presumably known) clustering evolution of the spectroscopic sample.  Thus, the  method requires that we have sufficient data to calculate the  clustering evolution of the spectroscopic sample, and assumes that the  bias in the photometric sample varies linearly with the bias in the  spectroscopic sample.  This assumption will break down as we include  measurements at small radii as the clustering moves in to the non-linear  regime.  
Finding the scales at which the bias evolution becomes too great for effective correction is the subject of this paper.

As in N08 and \citet[]{Matt:10} (see these references for the full derivation), we have a relation between the cross-correlation function of the spectroscopic and photometric samples and the normalized redshift distribution, $\phi(z)$ of the sample that we are attempting to estimate, given by:
\begin{equation}\label{eq:newman1}
w_{sp}(\theta,z) = \frac{\phi(z)H(\gamma)r_{\rm{0,sp}}^{\gamma_{sp}}\theta^{1-\gamma_{sp}}D_{\rm{A}}^{1-\gamma_{sp}}}{dl/dz}
\end{equation}

\noindent where $\gamma$ is the power law slope of the correlation function, $D_{\rm{A}}$ is the transverse comoving distance, $H(\gamma)=\Gamma(1/2)\Gamma[(\gamma-1)/2]/\Gamma(\gamma/2)$ and l(z) is the comoving distance to redshift z.  If we assume that the correlation function is a power law of the form $w_{sp}(\theta)=A_{sp}\,\theta^{1-\gamma}$ between \rmin~and \rmax~then our one bin measurement of the overdensity is proportional to the amplitude of the cross-correlation signal, $A_{sp}$.
 The unknown quantities in Equation~\ref{eq:newman1} are $\phi(z)$ and $r_{\rm{0,sp}}^{\gamma_{sp}}$.  We cannot evaluate $r_{0,sp}$ and $\gamma_{sp}$ as a function of redshift directly, as we do not know the redshift distribution of the ``photometric'' sample, though we can estimate them using other measured quantities.  Assuming that the galaxy bias is linear, we can estimate the cross-correlation parameters from power law fits to the autocorrelation functions of the spectroscopic and photometric samples, $\gamma_{sp}=(\gamma_{ss}+\gamma_{pp})/2$ and $r_{0,sp}^{\gamma_{sp}}=(r_{0,pp}^{\gamma_{pp}}r_{0,ss}^{\gamma_{ss}})^{1/2}$, where $\gamma_{ss}$ and $r_{0,ss}$ are measured from the projected correlation function, $w_{p}(r_{p})$ of the spectroscopic sample, and $\gamma_{pp}$ is the measured power law slope of the photometric sample angular autocorrelation function.  Rearranging Equation~\ref{eq:newman1}, we find:

\begin{equation}\label{eq:newman2}
\phi(z)=A_{sp}(z)\frac{dl/dz}{H(\gamma)r_{0,sp}^{\gamma}D_{\rm{A}}^{1-\gamma}}
\end{equation}
\noindent with the estimated $r_{0,sp}$ entering in the denominator, diminishing the effect of the bias.  The updated  $\phi(z)$ can now be used in Equation~\ref{eq:newman1} to obtain an  updated value for $r_{0,sp}$, and the process can be iterated until  convergence is reached.  Note that this iterative procedure simply  estimates a single, average value for $r_{0,pp}$ and assumes that  $r_{0,sp}$ scales linearly with $r_{0,ss}$ to improve the redshift  recovery.  If $r_{0,sp}$ is evolving in a non-linear fashion with  redshift this method will not correct the redshift distribution  appropriately.  The shape of the redshift distribution also impacts the effectiveness  of the iterative correction: as we are assuming that the clustering length of the photometric sample is proportional to that of the spectroscopic sample, the optimal iterative solution will work best at the  mean redshift of the sample.  For a compact and peaked redshift distribution the linear bias  assumption will be a good approximation of the true bias.  For a broad,  or multiply peaked distribution, deviations from the linear  approximation will become more problematic.  If  we can break the photometric sample into narrow  subsets in redshift it is possible to mitigate this problem.  This will be examined in Section~\ref{tomography}.

\subsection{Choice of $r_{min}$ and $r_{max}$}\label{rminrmax}
The photometric overdensity is measured over a constant range of projected physical scales around each spectroscopic object, bounded by an inner and an outer radius, \rmin~ and \rmax.  The choice of these radii affects the recovery in a number of ways and the values for \rmin~and \rmax~serve as the primary tuning parameters for the fidelity of the recovered redshift distribution.

For cases where the photometric catalog contains some fraction of the spectroscopic catalog a complication can arise.  Excess signal from the cross-matches between objects in the catalogs may boost the recovery signal if we allow for \rmin = 0, or if astrometric uncertainties lead to mismatch of spectroscopic and photometric objects.  For our simulations, we have explicity excluded such objects, so small radius matches are unaffected by such contamination.  More broadly, at small physical separations (below $\sim\,1$ Mpc) the clustering of objects will become stronger and increasingly non-linear.  This increased amplitude is a strong indicator that there is more information to be extracted from the cross-correlation, albeit at the cost of bending some of the linear assumptions described in \S\,\ref{bias}.  We will discuss this issue further in Section~\ref{future}.  For the outer boundary of the annulus, as \rmax~ increases, more physically associated galaxies will be included in the annulus, but so will an increasing number of unassociated background sources.  The clustering signal declines as radius increases, so increasing \rmax~can degrade the signal to noise ratio of the measurement.  The optimum $r_{max}$ will depend on both the clustering of the sample and the density of the photometric source catalog.   

\section{Simulated Galaxy Catalogs}\label{sims}

The redshift recovery procedure is sensitive to the evolution of galaxy bias.  In order to test this sensitivity
we employ two sets of simulations: mocks based on Millennium light cones with a limited field of view \citep[]{Sper:05,Cro:06} and larger area mocks based on LasDamas simulations (McBride et al, in prep). 
To cover a wide variety of possible scenarios we will examine four mock data sets: 
\begin{enumerate}
\item No galaxy bias evolution. 
\item Evolving galaxy bias as expected for a realistic, magnitude limited sample.
\item A magnitude limited selection with an additional stellar mass cut used to recover the magnitude limited sample, as might be expected if our spectroscopic catalog was a particular galaxy type.
\item A mixed case with constant bias at low redshift and a magnitude limited selection at higher redshift used to recover a sample with smooth bias evolution. Such a distribution might arise from multiple populations or complex selection criteria.
\end{enumerate}
\noindent
The first two cases have spectroscopic and photometric data drawn from the same underlying distribution, and thus identical galaxy bias properties.  Cases iii and iv have different bias properties for the spectroscopic and observed samples, as will be discussed in the following subsections.
We use the LasDamas simulations for the constant bias (item i) and mixed evolution samples (item iv), and the Millennium simulations for the magnitude selected and stellar mass selected samples (items ii \& iii ). 

\subsection{LasDamas Based Mock Catalogs}\label{damas}

The LasDamas catalogs used in this paper are a customized galaxy data set generated from the dark matter simulations of the LasDamas project (McBride et al, in prep)\footnote{We note that these mocks are {\it not} part of the ``publicly available'' mocks accessible from the LasDamas website { \texttt {http://lss.phy.vanderbilt.edu/lasdamas/}}}.  These galaxy mocks were constructed for testing this method, and do not explicitly fit to observed SDSS data, as is done in the full LasDamas simulations.  This enabled us to extend the redshift range beyond $z > 1$, with samples spanning $0.03\leq z\leq 1.33$, and covering a $9\times14$ degree patch of sky.  The galaxy mocks are constructed from a static redshift output of one of the four large LasDamas boxes (the \emph{Carmen} simulations)  at $z = 0.5$. We defined friends-of-friends halos \citep{fof} with a linking length of 20\% of the mean inter-particle separation. We assigned mock galaxies based on a simple $3$ parameter halo occupation distribution model \citep[i.e. HOD; ][]{berlind:02}.  To achieve the variable bias, the HOD is varied to reduce the number density as a function of redshift (thereby increasing the bias).  
The LasDamas simulations assume a flat \lcdm\ cosmology with $\Omega_{m}=0.25$, $\Omega_{\Lambda}=0.75$, $h=0.70$, and $\sigma_8 = 0.8$.

We construct two spectroscopic catalogs, populating the same dark matter halo catalogs with two different types of galaxy bias applied to generate the data:
\begin{itemize}
\item A constant bias for the whole redshift range consisting of approximately 120,000 galaxies over 125 deg$^{2}$
\item A constant bias over ($0.2\,\leq\,z\,\leq\,0.77$,) then mimicking an apparent magnitude limited selection for $z>0.77$ with about 235,000 galaxies over 500 deg$^{2}$.
\end{itemize}
\noindent
We refer to the first as the ``constant bias'' sample and the second as the ``mixed bias evolution'' sample.  For the photometric samples we create distributions drawn from the same constant bias case, as well as an additional dataset with bias is chosen such that the density decreases linearly with redshift over the range $0.2\,\leq\,z\,\leq1.33$ (referred to as the "linear density evolution" sample.  For these samples we create:
\begin{itemize}
\item A bimodal sample with galaxies in the ranges $0.4 < z < 0.6$ and $0.8 < z < 1.1$ containing about $350,000$ for the constant bias sample and about $592,000$ galaxies for the linear density evolution sample.
\item A Gaussian centered at $z=0.75$ and width $\sigma_{z}=0.10$ with $\sim 145,000$ for the constant bias case and $\sim 410,000$ galaxies for the linear density evolution sample.
\end{itemize}

\subsection{Millennium Galaxy Mock Catalogs}\label{millen}
The Millennium Simulation galaxy mock catalogs of \citet[]{Cro:06}\footnote{ Available at:\\ {\texttt{http://web.me.com/darrencroton/Homepage/SDSS-DEEP2.html}}} are light  cones populated with galaxies generated from semi-analytic models that follow the  prescriptions of ~\citet[]{Cro:06} and \citet[]{Kitz:07}.  We use the  four 2$\times$2 degree ``DLS'' cones, designed to match the footprints  of the Deep Lens Survey \citep[]{Witt:02}.  The light cones contain 17.4 million galaxies with redshifts spanning $0<z<3$ over 16  deg$^{2}$ with a magnitude limited r-band depth of $r=29.0$, which cover a redshift range large enough that significant galaxy bias evolution will occur.  Areal coverage is limited enough that sample variance will be a significant factor in some measurements.  The Millennium simulation assumes cosmological parameters $\Omega_{m}=0.25$, $\Omega_{\Lambda}=0.75$, $h=0.73$, and $\sigma_{8}=0.9$. 

We construct two ``spectroscopic'' catalogs with  known redshifts, and two ``photometric'' catalogs, where no redshift  information is retained.  For the spectroscopic sets:
\begin{itemize}
\item We randomly select  $\sim2\%$ ( approximately 325,000  galaxies) of the magnitude limited sample that will have a galaxy bias  evolution matching that of the underlying sample.  
\item We design a galaxy sample of  roughly the same size as  the previous sample (335,000 galaxies) which contains all galaxies with stellar mass greater than $2.3\times10^{10}\  M_{\sun}$.
\end{itemize}
\noindent
We refer to the first as the ``bias evolution'' sample and the second as the ``masscut'' sample.  Each has a surface density of approximately 5.6 galaxies/arcmin$^{2}$.
For the photometric samples, we draw from the same underlying magnitude limited distribution used in the evolving bias scenario, thus they have identical bias evolution properties to the evolving bias case.  We construct two samples from the simulation data:
\begin{itemize}
\item A bimodal distribution with 1.9 million galaxies covering the ranges $0.9 < z < 1.1$ and $1.9 < z < 2.1$.
\item A Gaussian distribution of about 690,000 galaxies centered at $z = 1.5$ with width $\sigma_z = 0.15$.
\end{itemize}
As will be shown in later sections, despite similar redshift spans and identical spectroscopic catalogs, the effects of evolving bias on the recovered redshift distribution for these two samples can be quite dissimilar. 
The stellar mass selection gives our spectroscopic sample different  bias properties than the photometric samples, enabling us to test the  efficacy of the recovery algorithm in the presence of stronger, non-representative bias.  The evolving bias scenario has identical bias in the spectroscopic and photometric samples.  The iterative procedure should perfectly recover the redshift distribution when the biases are the same, however, we will see that this is not the case when we measure the clustering length based on the large linear regime scales but use non-linear clustering information to reconstruct the distribution.

\subsection{Clustering Measurements}\label{clusterstuff}

The recovery procedure requires  fits to the projected correlation function of the spectroscopic datasets, as well as the  slope and amplitude of the two point autocorrelation functions of the photometric samples.
For the Millennium data sets the projected correlation functions of the spectroscopic samples are well fit by a power law form, and lack a strong 1-halo break.  As the slope of the power law shows little variation, we fix it at $\gamma_{ss}=1.8$ when fitting for the correlation length, $r_{0,ss}$ and use only projected separations greater than $r_{p}>300kpc$.  Thus, non-linear evolution in the one halo regime will not be reflected in the clustering lengths used in the iterative corrections.
We fit a parabolic form to the $r_{0,ss}$ data to smooth small scale redshift dependence induced by the measurement errors.
For the angular correlation function fits of the photometric samples, we measure best fit power law slopes of $\gamma_{pp}=1.75\pm0.13$ for bimodal and $\gamma_{pp}=1.68\pm0.11$ for the Gaussian distribution.  

For the LasDamas data, both the constant bias and mixed bias data sets show a strong one halo component and a break in their projected correlation functions, with slopes of $\gamma_{ss}=1.45$ for the constant bias data set and $\gamma_{ss}=1.8$ for the mixed bias case, nearly independent of redshift, at $r_{p}>300$ kpc beyond the one-halo break.  Once again we fit for $r_{0,ss}$ using only this "quasi-linear'' regime.


\begin{figure*}
   \centering
   \includegraphics[width=.99\hsize]{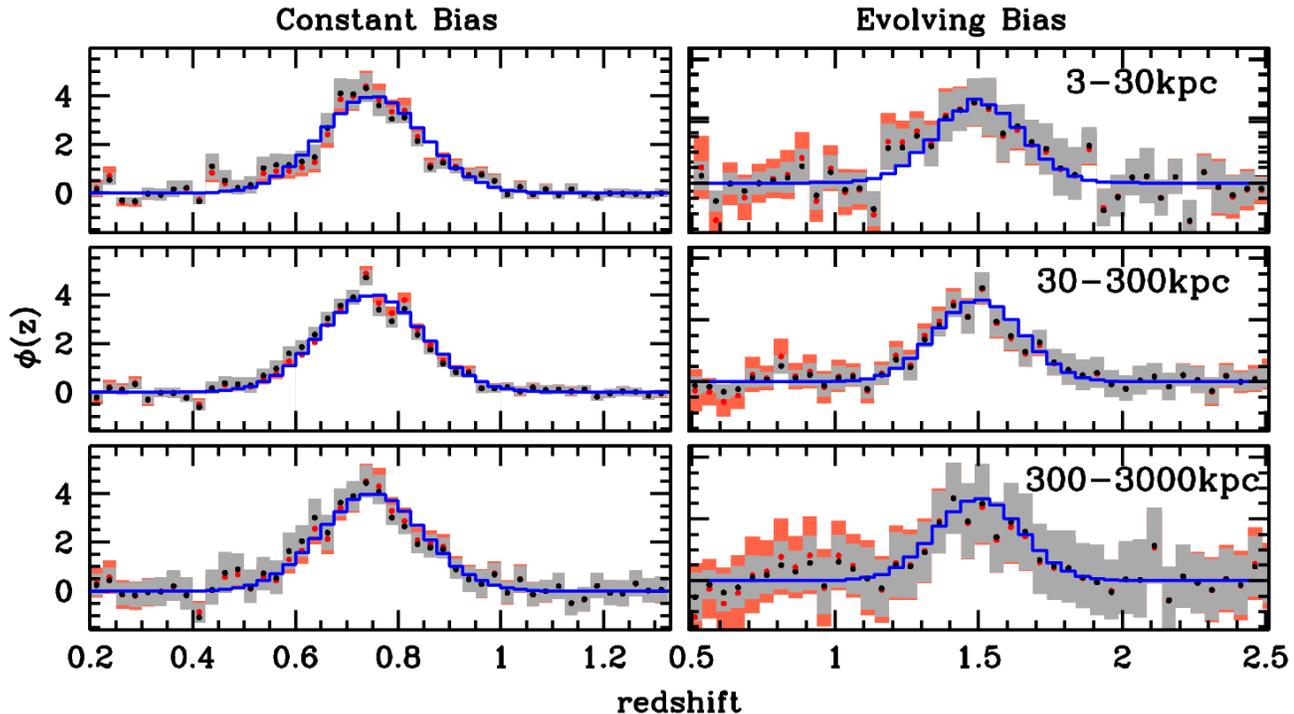}
   \caption{Recovered  Gaussian redshift distributions for the LasDamas constant bias (left)  and Millennium evolving bias (right) spectroscopic samples for three  decade width sets of \rmin~ and \rmax.  Red  points are the results before the iterative bias correction is applied,  while black points with gray errors are after the iteration.  The blue  histogram shows the actual redshift distribution of the photometric  sample.  The more centrally peaked distribution is less sensitive to  bias evolution than the broader bimodal distribution.\label{Gauss_plot}}
\end{figure*}

\begin{figure*}
   \centering
   \includegraphics[width=.99\hsize]{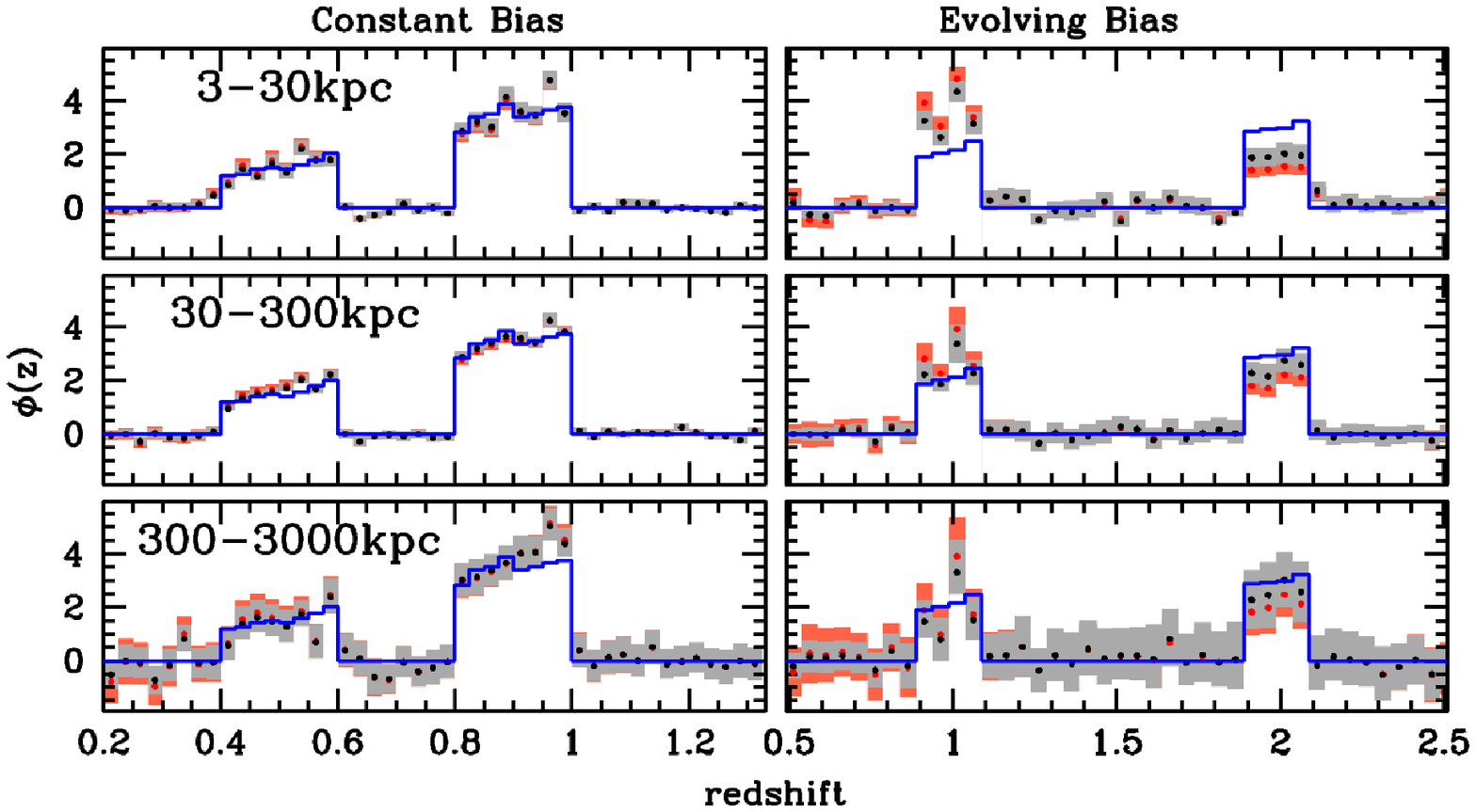}
   \caption{Recovered  bimodal redshift distributions for the LasDamas constant bias (left)  and Millennium evolving bias (right) spectroscopic samples for three  decade width sets of \rmin~ and \rmax.  Red  points are the results before the iterative bias correction is applied,  while black points with gray errors are after the iteration.  The blue  histogram shows the actual redshift distribution of the photometric  sample.  In the case of no bias evolution the recovery works well on all  scales, while evolving bias induces a skew in the recovered  distribution.\label{bimodal_plot}}
\end{figure*}

\section{Results}\label{results}
The effectiveness of cross-correlation methods in recovering redshift distributions is dependent on many factors.  Some of these (e.g. spectroscopic completeness or galaxy bias evolution) are determined by the survey data itself.  Others (the scale used for the cross-correlations, the redshift binning of the spectroscopic samples and weighting of the cross-correlation pairs) are nearly free parameters that can be used to tune the recovery.  Of these free parameters, the choice of scale has the most significant effect on the recovery due to its direct linkage to the galaxy bias dependence of the recovery.  In the N08 iterative technique, measurements are made on large enough scales (several Mpc) that the linear bias can be removed.  For the purposes of our analyis, we consider a much broader range of scales, from the linear to the quasi-linear ($\sim 1$ Mpc) down to the non-linear scales of a few kiloparsecs.  Our full analysis tests a wide range of scales, covering $3 \le$\rmin $\le 3000$ kpc and $10 \le$\rmax $\le 5000$ kpc, but for illustrative purposes we will show results only for three representative decade-width scales: $3 < r < 30$ kpc, $30 < r < 300$ kpc and $300 < r < 3000$ kpc.  

Before examining the redshift recoveries for these scales, a word about the smallest scales: Since we are using simulated data with perfect astrometry, we can distinguish perfectly between galaxies at scales where real data sets with noise from astrometric calibration and atmospheric blurring would likely struggle.  Applying these techniques to real data on those scales would likely mean that cross-contamination between the spectroscopic and photometric samples would dominate the recovered signal.  We have experimented with differing levels of cross-contamination between our simulated samples and find that the behavior of the recovered distributions is highly dependent on the choice of simulated spectroscopic sample.  To avoid this additional complication, we have chosen to eliminate all spectroscopic objects from our photometric catalogs and vice versa and defer further exploration of this effect to a future publication.

We estimate  errors on the redshift distributions with a spatial jackknife.  This consists of subdividing the sample into $N$ contiguous regions of the sky, each with approximately equal area.  We then perform each measurement $N$ times, each time leaving out one region.  We then estimate the jackknife variance as:
\begin{equation}
Var(x)=\frac{N-1}{N}\sum_{i=1}^{N} ({x_{i}-\bar{x}})^{2}
\end{equation}
\noindent where $x_{i}$ is the measurement for the ith region and $\bar{x}$ is the mean for the entire sample. 

\subsection{Recovery Scales and Populations}\label{scales}
In practial terms, several different bias scenarios may arise depending on the type of data selected.  We might select a population with very little expected bias evolution (e.~g.~ Luminous Red Galaxies), slowly evolving bias over a broad redshift interval (e.~g.~ field galaxies), or complex evolution due to the presence of multiple populations (e.~g.~a tomographic redshift bin with outliers).  For this reason we study the redshift reconstruction in several bias scenarios.  The shape of the redshift distribution of the photometric sample also plays a role: even a sample with strong galaxy bias evolution will not show significant relative bias change if the redshift interval of the recovery is sufficiently narrow.  Conversely, even slight bias evolution over a broad redshift interval may become significant when we include additional information from non-linear scales.  We test the recovery algorithm on two types of distributions to explore these effects: a centrally peaked Gaussian distribution and a bimodal distribution.

Figure~\ref{Gauss_plot} shows the recovered redshift distributions of the Gaussian photometric samples 
for both the LasDamas constant bias scenario and the Millennium evolving bias at our three representative scales.  Red points show the distribution before the iterative correction of Equations~\ref{eq:newman1} and~\ref{eq:newman2} is applied, while black points show the results after the correction.  In the constant bias case the iterative correction should have almost no effect, which is seen in the small difference between the pre- and post-iteration recoveries.  Interestingly, the method performs extremely well in the absence of bias evolution, down to the smallest scales, and including scales that span the break in the LasDamas correlation function. 

The centrally peaked Gaussian distribution, with most galaxies close to the mean redshift where our $r_{0,pp}$ estimate is most accurate, shows little sensitivity to effects of the evolving bias.  More compact and symmetric photometric distributions will be less affected by galaxy bias evolution, which enables us to push the recovery to smaller scales.  We will discuss this further in Sections~\ref{tomography} and~\ref{purity}.  For the constant bias scenario the best fit is $50 \le r \le 100$ kpc with a reduced $\chi^{2}=1.42$, though the change in $\chi^{2}$ is not particularly sensitive to the exact values of \rmin~ and \rmax (i.~e.~ the likelihood surface is very flat).

For the evolving bias the best fit occurs for $100 \le r \le 300$ kpc with a reduced $\chi^{2}=0.107$.  The $\chi^{2}$ value  significantly below 1.0 shows that we are overestimating our error bars  for the Millennium sample, which is not unexpected: with only 16 square  degrees available in the Millennium light cones we use only 32  jackknife samples to estimate the errors on fits for 99 bins.  This  appears to only affect the amplitude of the errors, and not the overall  structure of the covariance matrix.  The centrally peaked distribution shows little sensitivity to bias evolution even down to the smallest scales probed, and we accurately recover the distribution at all scales.

Figure~\ref{bimodal_plot} shows the recovered redshift distributions of the bimodal samples for the constant bias and evolving bias spectroscopic samples.    The best fit for the bimodal sample occurs at  $10 \le r \le 50$ kpc with a reduced $\chi^{2}$ of 1.16. 
The $\chi^{2}$ values  are similar for the bimodal and Gaussian distributions, showing that in  the absence of bias evolution the recovery performs accurately  regardless of the shape of the redshift distribution.  
The most notable feature in the evolving bias scenario is the relative amplitude of the two peaks.  The bimodal sample shows a clear bias before the iterative correction is applied, with larger discrepancies as the annulus moves to smaller physical scales.  This is as expected, since this bimodal configuration is particularly sensitive to bias evolution.  Because the iterative correction estimates a single, average value of $r_{0,pp}$ for the sample the bias correction is most accurate near the mean redshift of the photometric distribution.  The bimodal sample has a mean redshift of $z=1.59$, between the two peaks where no galaxies are located.  Also of note is the fact that even with identical bias evolution we introduce error into the recovered distributions even after iterative correction.  This is mainly due to the fact that we (purposely) measure the clustering length using only large ($>300\,kpc$) scales, and to a lesser extent due to the empirical estimation of the clustering length with finite samples that can also introduce errors.  We note, however, that the iterative technique does accurately recover the distributions when only the large scale clustering information is used, as expected.

The effectiveness of the iteration in correcting for the bias is obviously reduced as the radius of the annulus decreases, though errors due to covariance between bins also increase as \rmax~grows and more unassociated galaxies are included in the estimate.  
The best fit values are for intermediate scales, with a minumum at $200 \le r \le 300$ kpc and $\chi^{2}=0.334$.    The best fits at intermediate scales  balance the increasing influence of the bias at small scales with the concurrent increase in signal to noise and decreased bin to bin covariance that comes with smaller physical apertures.  It is clear that small scale information greatly increases bias in the recovery, and should not be used to recover broad redshift distributions.

Table~\ref{chi_table} lists the reduced $\chi^{2}$ values for the representative scale distributions both before and after the N08 iteration is applied.  The success of the iterative technique in aiding the recovery procedure is varied.  The iteration improves the recovery for every case in the evolving bias scenario, mixed results in the constant bias and mass cut scenarios, and mainly degrades results in the mixed/linear case.

\begin{table*}
\begin{center}
\caption{Reduced $\chi^{2}$ For Distributions before and after the N08 iterative correction}
\begin{tabular}{|c|p{1.5cm}|p{1.5cm}|p{1.5cm}|p{1.5cm}|}
\hline
Annulus & $\chi^{2}$/$N_{D}$ pre-iteration & $\chi^{2}$/$N_{D}$ post-iteration & $\chi^{2}$/$N_{D}$ pre-iteration & $\chi^{2}$/$N_{D}$ post-iteration\\
\hline
 & \multicolumn{2}{|c|}{Bimodal}  & \multicolumn{2}{|c|}{Gaussian} \\
\hline
   {\bf LD Constant Bias} \\
3-30 kpc & 3.34 & 3.11 & 1.55 & 1.59\\
30-300 kpc & 1.99 & 1.47 & 2.12 & 3.33\\
300-3000 kpc & 1.86 & 1.64 & 1.35 & 2.06\\
3-3000 kpc & 1.97 & 1.79 & 1.43 & 2.61\\ 
\hline
  {\bf Evolving Bias}  \\
3-30 kpc & 1.12 & 0.67 & 0.274 & 0.201\\ 
30-300 kpc & 0.566 & 0.258 & 0.229 & 0.148\\
300-3000 kpc & 0.400 & 0.216 & 0.264 & 0.215\\
3-3000 kpc & 0.496 & 0.270 & 0.237 & 0.207\\
\hline
  {\bf Mass Cut}  \\
3-30 kpc & 7.11 & 7.20 & 0.279 & 0.276\\
30-300 kpc & 1.99 & 2.04 & 0.156 & 0.155\\
300-3000 kpc & 0.381 & 0.376 & 0.215 & 0.204\\
3-3000 kpc & 0.743 & 0.783 & 0.514 & 0.521\\
\hline
  {\bf LD Mixed Bias}  \\
3-30 kpc & 228.2 & 513.1 & 366.08 & 122.8\\
30-300 kpc & 10.99 & 27.90 & 3.26 & 4.92\\
300-3000 kpc & 5.98 & 10.01 & 2.05 & 1.78\\
3-3000 kpc & 131.6 & 340.3 & 20.71 & 34.44\\

\label{chi_table}
\end{tabular}
\end{center} 
\end{table*}

\begin{figure}
   \centering
   \includegraphics[width=.99\hsize]{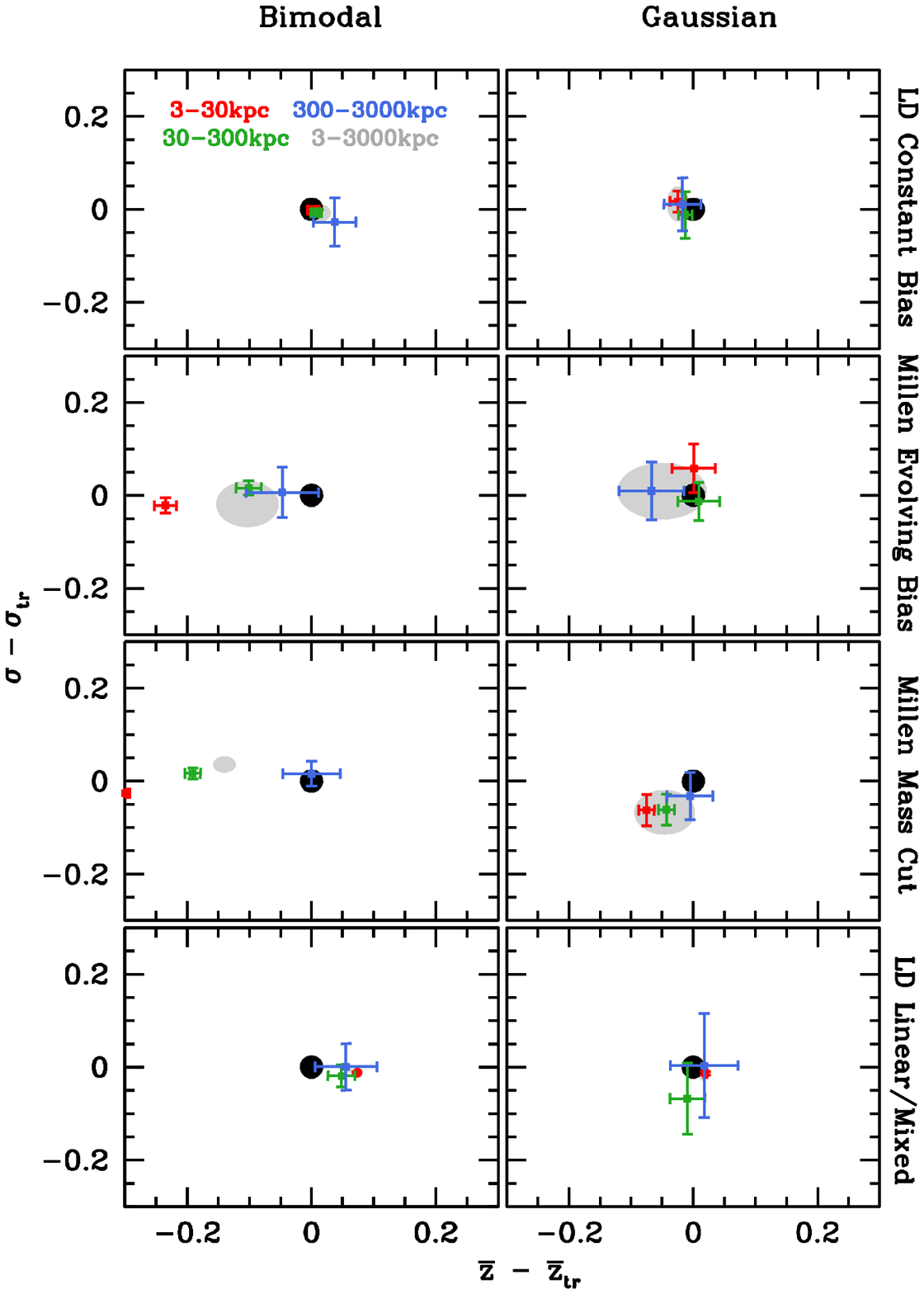}
   \caption{Measured deviation from the true redshift mean and width of the bimodal (left) and Gaussian (right) distributions for all four spectroscopic data sets. The truth is shown as the black dot, $3-30$ kpc (red), $30-300$ kpc (green), $300-3000$ kpc (blue), and $3-3000$ kpc (gray shaded) are shown for comparison.\label{all_moments}}
\end{figure}

For another more quantitative measure of the fidelity of the redshift recovery, we calculate the sample mean (average redshift) and standard deviation (square root of the sample variance, i.~e.~ the ``sample width'') for each distribution.  Figure~\ref{all_moments} shows the deviation from the true redshift mean $\bar{z}_{tr}$ and true sample width $\sigma_{tr}$ for the two cases shown in Figures~\ref{Gauss_plot} and~\ref{bimodal_plot}.  We show the three decade width bins and also results using information encompassing all three scales, with $3 \le r \le 3000$ kpc as a gray shaded ellipse.  As the information in each of the three annuli is independent, combining all scales should provide a higher signal-to-noise measurement of the redshift distribution.

The top two panels in Figure~\ref{all_moments} show that with  little or no expected bias evolution using all scales works extremely well at recovering the photometric sample distributions.
 For the evolving data set, the presence of  even modest bias evolution results in a misestimation of the mean redshift for the bimodel distribution, due to the relative amplitudes of the recovered peaks.
However, the sample width is largely  unaffected, as the lack of cross-correlation signal outside of the two bimodal bins provides a strong constraint on $\sigma$. 
In the centrally peaked Gaussian distribution the mean is more accurately recovered, but the uncertainty in the sample  width is increased.
This is due to the fact that the tails of the  distribution are now more affected by the difference in bias at low and  high redshift.

Unlike the top two rows, the bottom of Figure~\ref{all_moments} use simulated spectroscopic samples with very different bias profiles  from their respective photometric samples.
The stellar mass cut sample has stronger bias evolution than the magnitude limited sample that comprises the observed sample. 
The effects on the bimodel distribution in the evolving  bias scenario are exacerbated by the stronger bias in the mass cut  sample. Once again the mean redshift for the bimodal sample is  significantly skewed.

\begin{figure}
   \centering
   \includegraphics[width=.99\hsize]{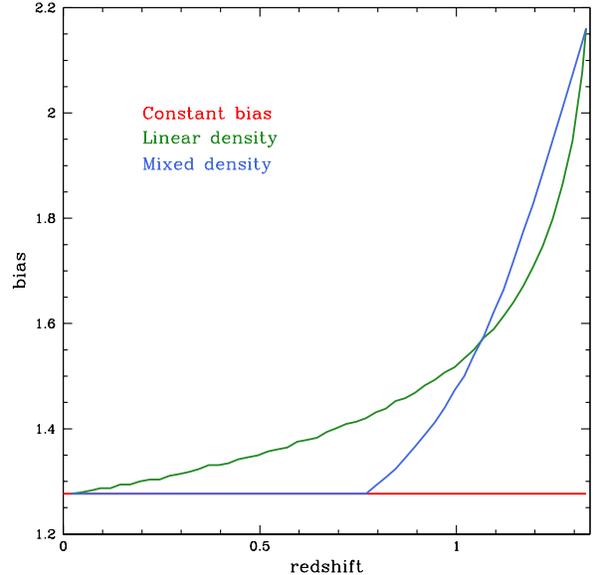}
   \caption{Linear galaxy bias as a function of redshift for the three LasDamas samples described in Section~\ref{damas}.  Red indicates the constant bias sample, green the sample with linear density evolution, and blue the sample with ``mixed'' bias evolution.  
.\label{biasfig}}
\end{figure}

To illustrate the differing bias of the three LasDamas samples we calculate the linear bias explicitly and show them in Figure~\ref{biasfig}.  The mixed/linear bias case uses the ``mixed'' bias data for the spectrscopic sample and the linear density sample for the photometric sample. 
In this case, the normal  tendency of the method to over-estimate signal at lower redshifts is  counteracted by the more rapid bias evolution in the spectroscopic sample, resulting in a  mean recovered redshift near the expected value for all scales.  
However, the difference in bias on the two  sides of the Gaussian distribution results in greatly increased scatter  in the recovered width.   Note that we did not have access to this information when computing the recovered distributions, in fact the bias is never explicitly calculated in the N08 iteration.  Instead, the clustering length of the spectroscopic sample, empirically measured from the correlation functions, is used to iteratively determine the best value for the photometric sample clustering length.

Overall, we see several trends: 
\begin{itemize}
\item{In absence of bias evolution, recovery S/N is always highest at the smallest scales.}
\item{In the presence of modest bias evolution,  intermediate scales ($\sim100 < r < 500$ kpc) offer the most reliable,  highest S/N recovery.}
\item{For extreme bias evolution, larger scales ($1000 < r < 3000$) offer the cleanest recovery.}
\item{For all cases, a centrally peaked redshift  distribution is far less sensitive to bias evolution, although outliers  can affect the recovered distribution width.}
\item{Small scale information should not be used to recover broad redshift distributions when the bias is known to evolve.}
\end{itemize}

\subsection{Tomographic Binning}\label{tomography}
The previous section discussed the recovery of broad redshift distributions.  However, most upcoming surveys will focus on determining the redshift distribution for narrow tomographic redshift bins for the purposes of measuring weak lensing and baryonic acoustic oscillations.  

The main limitation of the iterative method is that it relies on a single estimated value for $r_{0,pp}$.  If the clustering length of the photometric sample evolves differently than the spectroscopic sample, then the assumption that $r_{0,sp}$ scales as  $r_{0,sp}^{\gamma_{sp}}=(r_{0,pp}^{\gamma_{pp}}r_{0,ss}^{\gamma_{ss}})^{1/2}$ will not hold.  The iterative method essentially finds the best single value for $r_{0,pp}$ given the data.  However, if we can further subdivide our photometric sample in redshift, e.~g.~with some photometric redshift algorithm or color selection, we benefit in several ways:  First, we may now determine a best fit $r_{0,pp}$ over a smaller redshift range for each subsample, over which the bias presumably evolves less.  Second, having two values of $r_{0,pp}$ to estimate gives an additional free parameter.  Third, the signal to noise of the measurement increases, as by breaking our initial photometric dataset into multiple samples in redshift, we have removed a large number of physically unassociated galaxies from the correlation measurement that were adding to the background and diluting the signal.  

\begin{figure}
   \centering
   \includegraphics[width=.99\hsize]{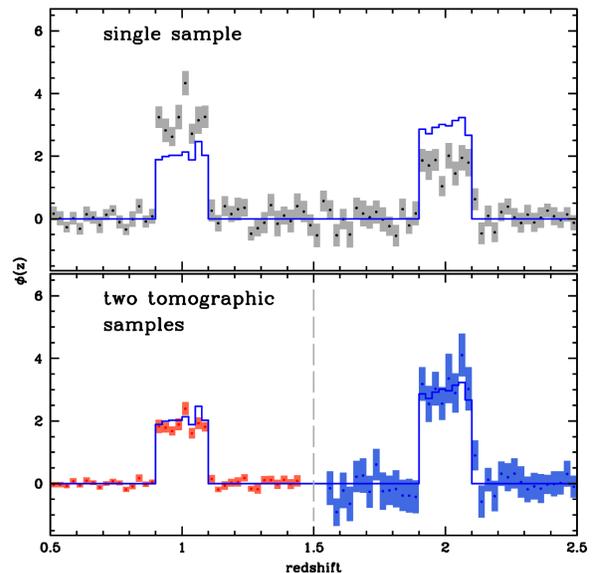}
   \caption{Top: Recovered redshift distribution for the bimodal Millennium light cone sample for an annulus of $3-30kpc$. Bottom: the same sample split into two redshift bins (overlapping points omitted for clarity).  The bottom panel shows that the amplitudes of the bimodal recovered low redshift (red) and high redshift (blue) samples are significantly less biased than the union of the two samples recovered at once.\label{millen_tomog}}
\end{figure}

Figure~\ref{millen_tomog} shows the result of splitting the bimodal sample for the evolving bias dataset (the same as shown in the top right panel of Figure~\ref{bimodal_plot}) into two distinct redshift bins and computing the recovered distributions for each individual bin.  This is done for \rmin~ and \rmax~ values of 3 kpc and 30 kpc, far into the non-linear regime and at much smaller scales than the best fit in the previous section.  The red points show the recovery for the low redshift bin, while the blue points show the recovery for the high redshift bin (overlapping points have been omitted for clarity).  The top panel shows the results of the same reconstruction using a single photometric sample.  The change in the size of the errors on the two individual recoveries is related to the normalization factor enabled by the two independent estimates of the best fit clustering length, which boosts/lowers the signal-to-noise of the low/high redshift recoveries.  In general, the more narrow the redshift range can be restricted, the smaller the optimum recovery scale will be, which results in an increase in S/N.   However, the presence of catastrophic outliers in certain photometric redshift ranges may be a concern when applying tomographic selections.  We address this in the following Section.

\subsection{Redshift Outlier Detection}\label{purity}
Selection of tomographic redshift bins in cosmological analyses, for instance with color cuts or photometric redshift cuts, can include data sets where degeneracies exist that include an unrelated population far outside the intended redshift range.  The most prominent example in optical photometric surveys is the common Lyman/Balmer break degeneracy, where low redshift ($z \sim 0.2-0.3$) blue galaxies are mistaken for very high redshift ($z \sim 2-3$) blue galaxies and vice-versa, due to their similar optical colors.  Such ``catastrophic outlier'' populations often result in two bimodal peaks widely separated in redshift, which we have shown (\S\,\ref{scales}) can be problematic for accurate redshift recovery.  However, we can use small scale information to diagnose the presence of such outlier populations.

In nearly all cases examined previously, reconstructing the redshift distribution at smaller physical scales results in smaller uncertainties, albeit at the expense of increased bias sensitivity.  This is expected, given the power law form of the correlation function we expect more signal on smaller scales.  We also note that the method does an excellent job at returning a null signal in areas where there is no overlap between the spectroscopic  and photometric samples, e.~g.~we see signal consistent with zero and small error bars outside the bimodal bins in Figure~\ref{bimodal_plot}.  

\begin{figure}
   \centering
   \includegraphics[width=.99\hsize]{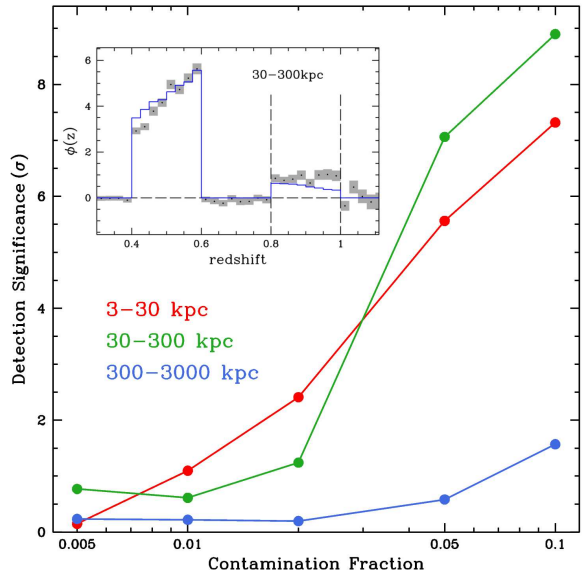}
   \caption{Detection significance of the secondary peak as a function of contamination fraction using the LasDamas mixed bias data set for three annuli.  The inset shows an example recovery with 10\% of the galaxies in the second peak.  The higher S/N per bin for smaller annuli enables us to detect contaminating objects at much greater significance than when using large scale information alone.
\label{contamfig}}
\end{figure}

We can use these features to test for the presence of interlopers, as done in \citet[]{Mor:12}, where the authors cross-correlate a high redshift luminous blue galaxy sample with spectroscopically confirmed galaxies to test for the presence of intermediate redshift elliptical galaxies with similar expected colors.  
We construct several data sets to test the influence of recovery scales on sensitivity to outlier populations.
Using the LasDamas mixed bias evolution data set, we construct samples where we  have a primary peak at the redshift of interest ($0.4\le z \le 0.6$) and a secondary peak due to color degeneracies ($0.8\,\leq\,z\,\leq\,1.0$) that contains between 0.5\% and 10\% of the total number of galaxies.  Figure~\ref{contamfig} shows detection significance (in terms of $\sigma$ determined from the $\chi^{2}$) as a function of contamination fraction.  The inset shows one recovered distribution as an example, with 10\% contamination and using an annulus of 30 $\le r \le$ 300 kpc.

Using the $300 \le r \le 3000$ kpc annulus we cannot reliably detect the secondary peak, however at smaller scales we see nearly all bins outside of the two peaks consistent with zero, and clearly detect non-zero signal in the range $0.8\,\leq\,z\,\leq\,1.0$ for contamination fractions above 2\%.
The ability to detect secondary peaks will depend on both the redshift evolution of the bias and the amount of separation between the two peaks in redshift space.  The influence of the bias evolution when using small scale information can cause us to misestimate the overall contamination fraction, though detection of any contaminants at all may be the goal.  While the recovery method does not directly inform us of which galaxies are degenerate, we can use the method to tailor photometric redshift cuts that lead to maximum purity in the sample by testing variations of the cuts and choosing those that minimize sample contamination.

\subsection{Alternative Bias Removal Technique}\label{approx}
  The application of the full iterative procedure discussed so far requires calculation of the photometric sample angular autocorrelation functions.  In actual surveys, complex selection and masking often make estimation of the correlation functions difficult.  We can simplify our analysis by, instead of assuming a linear relation between the spectroscopic and photometric samples, assume that the two samples have the same bias as calculated from $r_{0,ss}$ (or measurements from the literature).  We take the estimates of $r_{0,ss}$ estimated for $r_{p}\,>\,300kpc$ discussed in in Section~\ref{millen} and calculate the bias evolution of the spectroscopic sample as a function of redshift.  In place of the full iterative procedure, we then simply divide our initial estimate of $\phi(z)$ by this relative bias and renormalize.  

\begin{figure}
   \centering
   \includegraphics[width=.99\hsize]{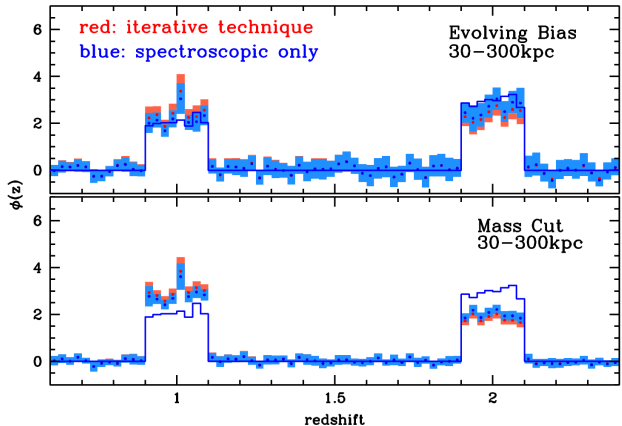}
   \caption{Comparing the iterative and approximate bias removal techniques for 30 $\le r \le$ 300 kpc.  The top panel shows the recovered distributions for the Millennium sample with evolving bias using the full iterative technique (red), and using only the spectroscopic bias (blue).  The bottom panel shows the same recovery using the stellar mass selected spectroscopic sample.  The simple bias approximation can provide as good or better estimates of the redshift distribution when the bias evolution of the two samples is similar.   
\label{approxbias}}
\end{figure}

The top panel of Figure~\ref{approxbias} shows a comparison between the initial estimate (black), the final iterative correction (red), and this alternative bias removal (blue) for the Millennium light cone simulation with \rmin$\,=\,30kpc$ and \rmax$\,=\,300kpc$ (though the conclusions hold at both smaller and larger scales as well).  The simple bias correction actually outperforms the iterative solution, with a $\chi^{2}=0.40$, compared to $\chi^{2}=0.61$ for the iterative method.  In retrospect, this is not unexpected: The photometric samples from the Millennium simulation used in Figure~\ref{approxbias} were drawn from the same underlying population as the spectrocopic sample, and thus have the same galaxy bias properties.  The linear bias approximation used in the iterative correction, calculating the correlation length $r_{0,sp}$ as the geometric mean of the spectroscopic correlation length and a single, constant value for the average photometric correlation length, actually lessens the predicted redshift evolution of the bias, particularly when using very wide redshift baseline for the photometric sample.  This is related to the improvements gained from splitting the sample into subsets in redshift, where we gain both in a smaller relative evolution in bias over the shorter redshift interval, and in the ability to estimate multiple values of $r_{0,pp}$ in the different redshift intervals.

Applying a stellar mass selection to the spectroscopic sample will change the galaxy bias evolution properties.  The bottom panel of Figure~\ref{approxbias} shows the recovery of the same photometric sample using the mass selected spectroscopic sample and corresponding bias estimate.  While the difference is less pronounced, using the spectroscopic bias to correct the amplitude again outperforms the iterative method, with $\chi^{2}=3.4$ versus the $\chi^{2}=5.1$ for the iterative method.  In practical terms, removing the bias evolution of the estimated sample using an estimate based solely on the spectroscopic sample can provide results competitive or better than those obtained from the full iterative technique.

\section{Discussion and Future Work}\label{future}
In this paper we have presented a study of a simple but powerful redshift recovery algorithm applied to realistic mock datasets, testing the inclusion of information from the non-linear clustering regime.  Our galaxy density estimator is equivalent to a one bin measurement of the cross-correlation function between user adjustable physical scales of \rmin~and \rmax.
We have shown that non-linear scales contain a wealth of information that can be exploited to increase S/N in the determination of redshift distributions compared to using only large scale information, and that the iterative technique used to mitigate the effects of bias evolution works well beyond the linear regime used to date for narrow distributions.  
Due to the wide variety of bias scenarios that may be present in real data we are limited to somewhat qualitative assessments.  However, these general conclusions are informative in future applications of the method.

We successfully recover the redshift distributions for several evolving and non-evolving galaxy bias configurations.  However, the non-linear biasing does incur increasing amounts of error as we push to smaller and smaller radii.  The optimum scale depends on the details of the photometric dataset, both in terms of bias properties and extent in redshift: narrow redshift distributions and those with little expected bias evolution can exploit clustering signal well into the non-linear regime, while broad redshift distributions or complex galaxy bias should be restricted to the more conservative limits at larger scales.   Furthermore, in our one bin treatment, larger values of \rmax~ lead to increasing covariance between redshift recovery bins, which increases the relative error when using large scales.  One must find the balance between the increased signal-to-noise and the accompanying increased sensitivity to galaxy bias when performing the recovery.

The iterative correction suggested by N08 and \citet[]{Matt:10} and employed in this paper has limitations. The assumption that the bias of the photometric sample scales linearly with the spectroscopic sample allows us to determine only a single value for the average cluster scaling between the spectroscopic and photometric samples via equation~\ref{eq:newman2}.  The technique works well at correcting for galaxy bias when used for large scales, but begins to fail, as expected, as the non-linear information is included.  We explored using an approximation of simply dividing by the bias of the spectroscopic sample in section~\ref{approx}, and found that this works well in many cases, though the same caveats that apply to the use of the iterative corrections apply.
The iterative correction works best at the mean redshift of the photometric sample, thus narrow redshift distributions peaked near the mean redshift are recovered much more accurately than broad distributions, as illustrated by the relative performance of the Gaussian and bimodal samples shown in Figures~\ref{Gauss_plot} and~\ref{bimodal_plot}.  If we are able to subdivide the distribution we wish to recover into narrower redshift ranges then we can recover the distribution more accurately, as the bias should evolve less over the smaller redshift interval (assuming a smoothly varying bias evolution).  This was illustrated in the simple example of breaking one of our bimodal samples into two bins in Section\,\ref{tomography}.  This is in line with the direction of the large future surveys (DES, LSST, etc...), where the strategy for determining cosmological parameters hinges on precisely determining the redshift distribution for a number of relatively narrow tomographic photometric redshift bins.  The tomographic bins planned for such surveys are ideal samples for including non-linear information in redshift recovery.  However, extra care will have to be taken if bins include any ``catastrophic outlier'' galaxies, where photometric redshift degeneracies cause some portion of the sample to lie at very different redshifts than that targeted by the selection. Such distributions will be susceptible to biasing, particularly when including information from non-linear scales.  In such cases, reverting to large \rmin~ values may be necessary.  Even in such cases, the non-linear regime can be used to accurately assess the presence or absence of catastropic outliers in the sample, as illustrated in the tests of sample contamination discussed in section~\ref{purity}.  We plan to carry out tests on more realistic tomographic photometric redshift bins based on improved simulations in an upcoming paper.

The method could be further improved by extending beyond the simple one-bin treatment used in this paper, particularly in cases where there is obvious non-power law form to the correlation functions, or where the slope of the power law changes substantially.  For example, in the determination of $r_{0,ss}$ we used only information at scales greater than 300 kpc, beyond the break in the correlation function, even when testing the recovery at the smallest scales.
An explicit fit to both the one halo and two halo portions of the correlation function would enable a more precise recovery.  However, the non-linear relation between galaxies and underlying dark matter at small scales will still leave the method susceptible to the influence of galaxy bias evolution. 

Having shown that the methods discussed in this paper can accurately recover redshift distributions using small scale clustering information, we will follow up with analyses using real data sets for both known and wholly novel redshift distributions (M\'{e}nard et.~al, in preparation).  This powerful technique will be an important and useful tool for both current and future photometric surveys.

\section*{Acknowledgements}

This work was supported by NSF Grant AST-1009514. Brice M´enard is supported by the NSF and the Alfred P. Sloan foundation.  
We thank the anonymous referee for suggestions that improved the content of the paper.
We used mock catalogs based on the LasDamas project; we thank the LasDamas collaboration for providing us with this data.
We thank Darren Croton for making the Millennium Simulation light cones used in this work publicly available.

\appendix

\section{Additional Recovery Parameters}\label{recov_params}
While the choice of the physical annulus defined by \rmin~and \rmax~is the dominant factor in determining the redshift recovery, we have the freedom to choose both an additional radial weighting of our pixelized aperture and the bin width of the spectroscopic sample.

\subsection{Annulus Weighting}\label{weight} 
\begin{figure}
   \centering
   \includegraphics[width=.99\hsize]{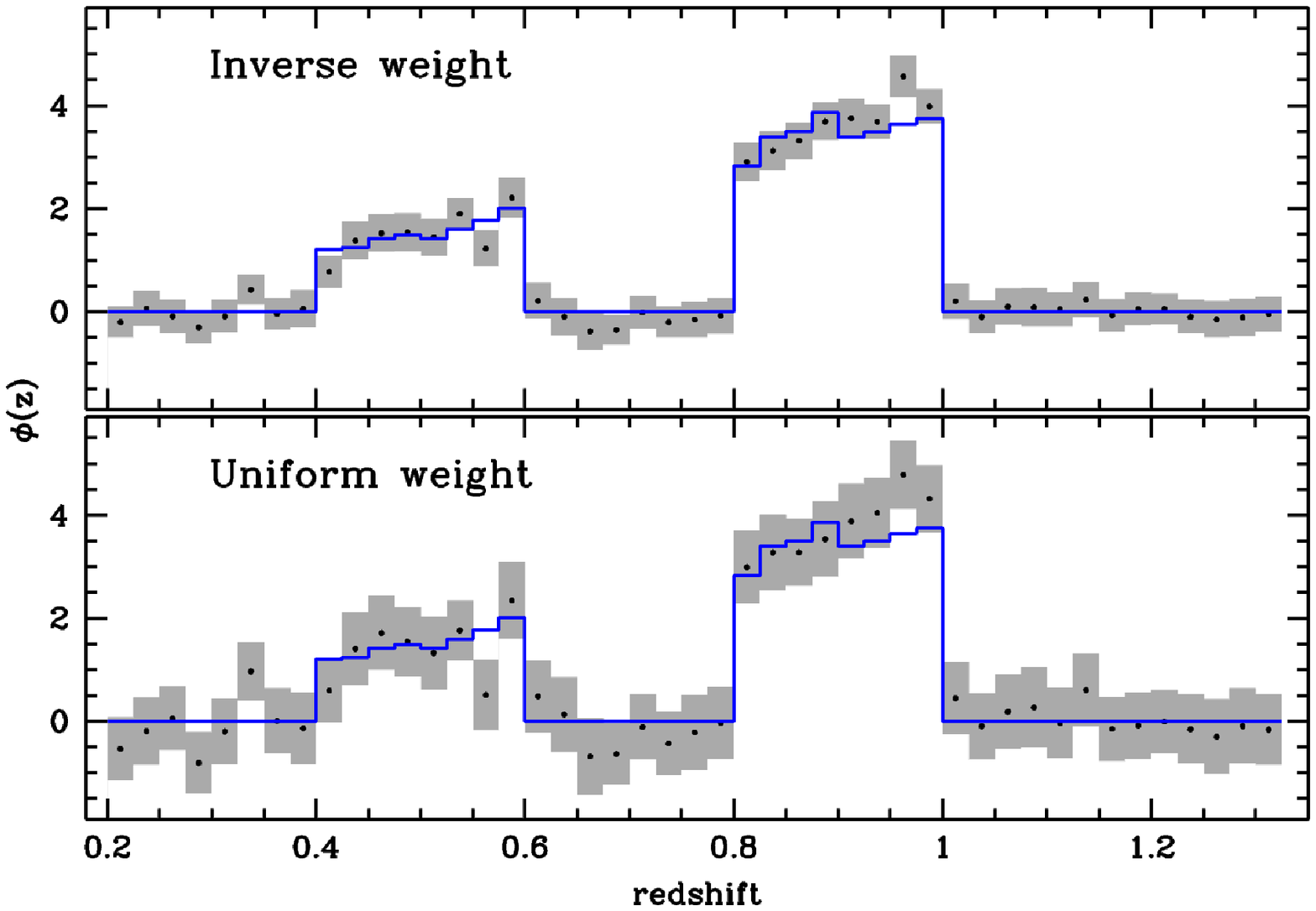}
   \caption{Recovered redshift distribution for bimodal sample of galaxies for the constant bias sample for the ``uniform'' density weight (left) and ``inverse'' density weight (right).  The magenta histogram shows the actual redshift distribution of the photometric sample.  The inverse weighting produces smaller error estimates, but is more sensitive to the effects of non-linear bias evolution.\label{weight_bimod}}
\end{figure}

The power law  form of the correlation  function shows that there is an increasing  amount of clustering  information at smaller angular and physical scales,  however there are  also fewer galaxies due to the decreasing area of the  annulus.  Similarly, while larger apertures decrease shot noise in  density  estimates, they also increase the number of unassociated  galaxies that  are included due solely to line of sight projection.  In  addition to a ``uniform'' density estimator, where we simply divide the  number of  galaxies within the pixelized annulus by the area in physical  units of  Mpc, we test an ``inverse'' weighted density  estimate,  calculating the density in the pixelized annulus and  weighting the density  in each pixel by the inverse of the distance from  the spectroscopic  object.  
Errors  are a combination of variations due to  large-scale-structure (which  becomes a more serious issue for surveys  with small areal footprint),  and Poisson shot noise.  We estimate errors  on the recovery empirically  with a spatial jackknife, which captures  both sources of error.  
For  illustration,  Figure~\ref{weight_bimod} shows the uniform vs.~inverse  weighting  recovered redshift distributions of the constant bias bimodal  sample  with a large outer radius of \rmax=3000kpc.  
There  is  a clear reduction of error when using the inverse weighting, which  is  observed at nearly all scales and in all samples tested, thus we employ  this inverse weighted estimator throughout the paper.  However,  because  the smaller scales are now weighted more heavily, this estimator is  more  sensitive to evolving galaxy bias.  
Therefore,  caution should be used when using the inverse weighting at very small  scales when galaxy bias evolution is known to be large.

\subsection{Redshift Binning}\label{binning}
The choice of binning for the spectroscopic sample is an additional free parameter that we must choose.  To construct our redshift distribution we take each spectroscopic galaxy and estimate the density of sources within the physical aperture defined by \rmin~and \rmax.  Then, we bin all spectroscopic objects within a redshift interval $\Delta\,z$ and take the mean of the density estimates within each bin to determine the amplitude of the redshift distribution estimate.  

Several factors influence the uncertainties resulting from a specific choice of redshift binning.  Errors are a combination of Poisson fluctuations, i.~e.~the number of spectrocopic galaxies included in the bin, sample variance, and the fractional error in the amplitude of the cross-correlation function.  The sample variance is fixed by large scale structure and the areal coverage of the survey.  The amplitude of the cross-correlation function depends on the width of the redshift bin, as using broader redshift bins lowers the amplitude of the cross-correlation signal.  Narrow redshift bins lead to a stronger cross correlation signal, however this must be balanced with Poisson noise from small samples within the bin.   In practice, the total signal-to-noise is not a strong function of bin-width choice for small bins.  However, the total signal-to-noise is significantly lower when using a small number of very broad bins.  Using a small number of bins effectively throws out information unnecessarily.

\begin{figure}
   \centering
   \includegraphics[width=.99\hsize]{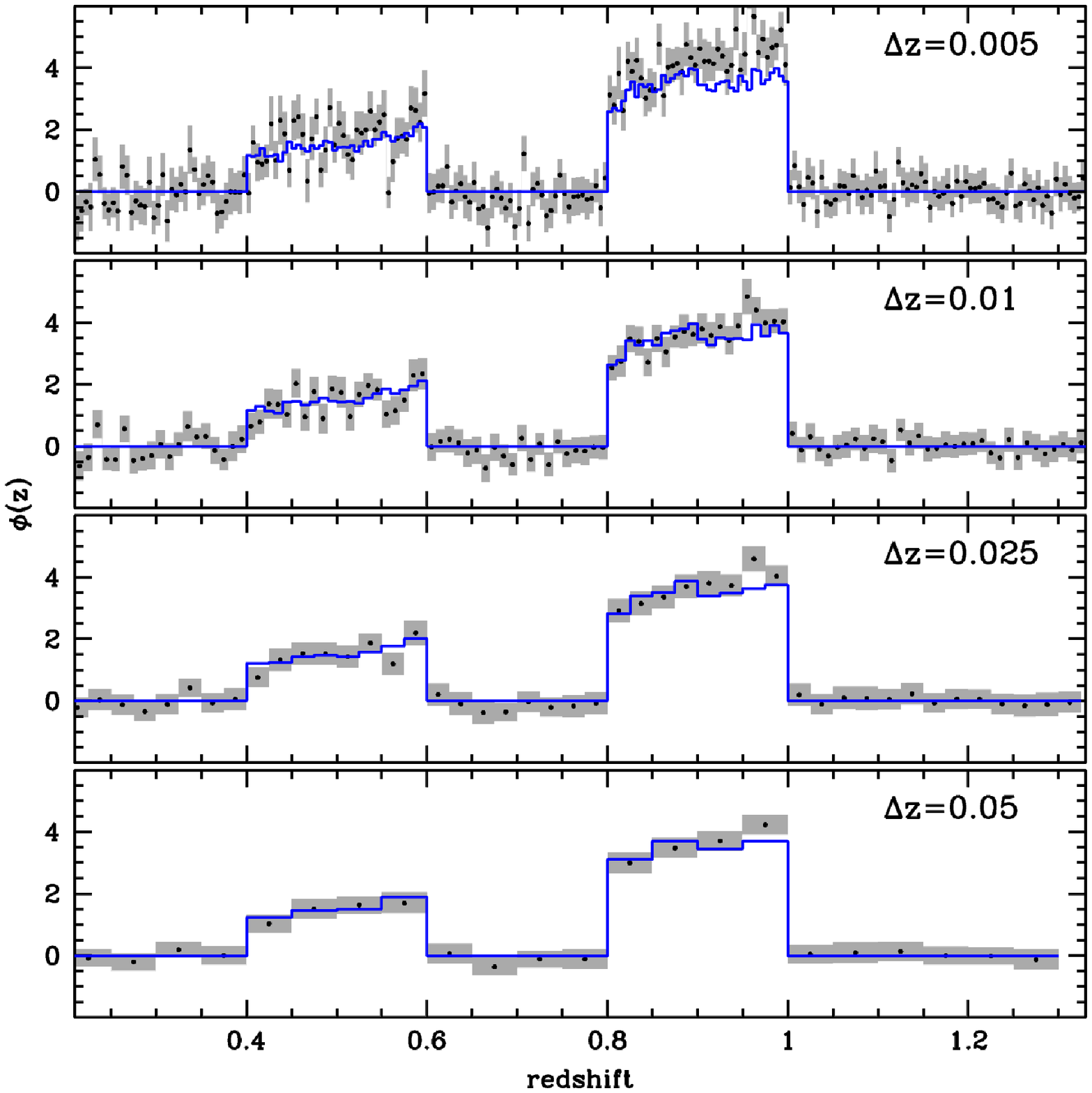}
   \caption{The effect of spectroscopic bin size on the recovered redshift distribution in a data set with no bias evolution.  Errors are a combination of large scale structure fluctuations and Poisson noise in the average density estimate.  The effect of Poisson fluctuations can be seen as the number of galaxies per bin decreases at $\Delta\,z=0.005$. \label{binsize}}
\end{figure}

Figure~\ref{binsize} shows the jackknife error estimates for several bin size choices using the LasDamas based mock dataset with no bias evolution and $30 \le r \le 300$ kpc.  
For the constant bias sample used in Figure~\ref{binsize} the optimal scale occurs at $\Delta\,z\approx\,0.005$ with $\sim200-800$ galaxies per redshift bin used to determine the mean density.  We expect adjacent bins to be increasingly correlated as $\Delta\,z$ decreases, as shared large scale structure near the bin boundaries should become more important.

All bins are correlated with each other, as expected, since the density estimate of background galaxies samples the distribution over the entire projected redshift range, with many galaxies falling within the physical annulus surrounding a spectroscopic object multiple times.  
This leads to a distinct correlation matrix structure: a strong diagonal and all off diagonal elements correlated at a similar ``floor'' level, the amplitude of which is determined by the size of the annulus and the width of the recovery.
A redshift bin of width $\Delta\,z=0.005$ corresponds to $\sim10-20$Mpc in comoving distance for $0.2\leq\,z\,\leq\,1.33$ probed in the recovery, much larger than the weighted physical distance, so the correlation matrix shows a strong diagonal component, but adjacent bins do not show excess correlation compared to widely separated bins.  The projected nature of the measurement leads to highly correlated bins, and the full covariance matrix must be used for proper error estimation.

\bibliographystyle{mn2e}
\bibliography{paper_v3.bib}

\begin{thebibliography}{26}
\expandafter\ifx\csname natexlab\endcsname\relax\def\natexlab#1{#1}\fi

\bibitem[{{Abazajian} {et~al}\mbox{.}(2009){Abazajian}, {Adelman-McCarthy},
  {Ag{\"u}eros}, {Allam}, {Allende Prieto}, {An}, {Anderson}, {Anderson},
  {Annis}, {Bahcall}, \& et~al.}]{sdss}
{Abazajian} K.~N. {et~al.}, 2009, \apjs, 182, 543

\bibitem[{{Ben{\'{\i}}tez}(2000)}]{Ben:00}
{Ben{\'{\i}}tez} N., 2000, \apj, 536, 571

\bibitem[{{Benjamin} {et~al}\mbox{.}(2010){Benjamin}, {van Waerbeke},
  {M{\'e}nard}, \& {Kilbinger}}]{Ben:10}
{Benjamin} J., {van Waerbeke} L., {M{\'e}nard} B., {Kilbinger} M., 2010,
  \mnras, 408, 1168

\bibitem[{{Berlind} \& {Weinberg}(2002)}]{berlind:02}
{Berlind} A.~A., {Weinberg} D.~H., 2002, \apj, 575, 587

\bibitem[{{Connolly} {et~al}\mbox{.}(1995){Connolly}, {Csabai}, {Szalay},
  {Koo}, {Kron}, \& {Munn}}]{Con:95}
{Connolly} A.~J., {Csabai} I., {Szalay} A.~S., {Koo} D.~C., {Kron} R.~G.,
  {Munn} J.~A., 1995, \aj, 110, 2655

\bibitem[{{Croton} {et~al}\mbox{.}(2006){Croton}, {Springel}, {White}, {De
  Lucia}, {Frenk}, {Gao}, {Jenkins}, {Kauffmann}, {Navarro}, \&
  {Yoshida}}]{Cro:06}
{Croton} D.~J. {et~al.}, 2006, \mnras, 365, 11

\bibitem[{{Cunha} {et~al}\mbox{.}(2012){Cunha}, {Huterer}, {Busha}, \&
  {Wechsler}}]{Cun:12}
{Cunha} C.~E., {Huterer} D., {Busha} M.~T., {Wechsler} R.~H., 2012, \mnras,
  2844

\bibitem[{{Cunha} {et~al}\mbox{.}(2009){Cunha}, {Lima}, {Oyaizu}, {Frieman}, \&
  {Lin}}]{Cun:09}
{Cunha} C.~E., {Lima} M., {Oyaizu} H., {Frieman} J., {Lin} H., 2009, \mnras,
  396, 2379

\bibitem[{{Davis} {et~al}\mbox{.}(1985){Davis}, {Efstathiou}, {Frenk}, \&
  {White}}]{fof}
{Davis} M., {Efstathiou} G., {Frenk} C.~S., {White} S.~D.~M., 1985, \apj, 292,
  371

\bibitem[{{Kerscher} {et~al}\mbox{.}(2000){Kerscher}, {Szapudi}, \&
  {Szalay}}]{Ker:00}
{Kerscher} M., {Szapudi} I., {Szalay} A.~S., 2000, \apjl, 535, L13

\bibitem[{{Kitzbichler} \& {White}(2007)}]{Kitz:07}
{Kitzbichler} M.~G., {White} S.~D.~M., 2007, \mnras, 376, 2

\bibitem[{{Ma} {et~al}\mbox{.}(2006){Ma}, {Hu}, \& {Huterer}}]{Ma:06}
{Ma} Z., {Hu} W., {Huterer} D., 2006, \apj, 636, 21

\bibitem[{{Matthews} \& {Newman}(2010)}]{Matt:10}
{Matthews} D.~J., {Newman} J.~A., 2010, \apj, 721, 456

\bibitem[{{Matthews} \& {Newman}(2012)}]{Matt:11}
{Matthews} D.~J., {Newman} J.~A., 2012, \apj, 745, 180

\bibitem[{{Mitchell-Wynne} {et~al}\mbox{.}(2012){Mitchell-Wynne}, {Cooray},
  {Gong}, {Bethermin}, {Bock}, {Franceschini}, {Glenn}, {Griffin}, {Halpern},
  {Marchetti}, {Oliver}, {Page}, {Perez-Fournon}, {Schulz}, {Scott}, {Smidt},
  {Smith}, {Vaccari}, {Vigroux}, {Wang}, {Wardlow}, \& {Zemcov}}]{Mit:12}
{Mitchell-Wynne} K. {et~al.}, 2012, ArXiv e-prints

\bibitem[{{Morrison} {et~al}\mbox{.}(2012){Morrison}, {Scranton}, {M{\'e}nard},
  {Schmidt}, {Tyson}, {Ryan}, {Choi}, \& {Wittman}}]{Mor:12}
{Morrison} C.~B., {Scranton} R., {M{\'e}nard} B., {Schmidt} S.~J., {Tyson}
  J.~A., {Ryan} R., {Choi} A., {Wittman} D., 2012, ArXiv e-prints

\bibitem[{{Newman}(2008)}]{New:08}
{Newman} J.~A., 2008, \apj, 684, 88

\bibitem[{{Nikoloudakis} {et~al}\mbox{.}(2012){Nikoloudakis}, {Shanks}, \&
  {Sawangwit}}]{Nik:12}
{Nikoloudakis} N., {Shanks} T., {Sawangwit} U., 2012, ArXiv e-prints

\bibitem[{{Quadri} \& {Williams}(2010)}]{Qua:10}
{Quadri} R.~F., {Williams} R.~J., 2010, \apj, 725, 794

\bibitem[{{Roberts} \& {Odell}(1979)}]{Rob:79}
{Roberts} D.~H., {Odell} S.~L., 1979, \aap, 76, 254

\bibitem[{{Schneider} {et~al}\mbox{.}(2006){Schneider}, {Knox}, {Zhan}, \&
  {Connolly}}]{Sch:06}
{Schneider} M., {Knox} L., {Zhan} H., {Connolly} A., 2006, \apj, 651, 14

\bibitem[{{Schulz}(2010)}]{Sch:10}
{Schulz} A.~E., 2010, \apj, 724, 1305

\bibitem[{{Seldner} \& {Peebles}(1979)}]{Seld:79}
{Seldner} M., {Peebles} P.~J.~E., 1979, \apj, 227, 30

\bibitem[{{Springel} {et~al}\mbox{.}(2005){Springel}, {White}, {Jenkins},
  {Frenk}, {Yoshida}, {Gao}, {Navarro}, {Thacker}, {Croton}, {Helly},
  {Peacock}, {Cole}, {Thomas}, {Couchman}, {Evrard}, {Colberg}, \&
  {Pearce}}]{Sper:05}
{Springel} V. {et~al.}, 2005, \nat, 435, 629

\bibitem[{{Wittman} {et~al}\mbox{.}(2002){Wittman}, {Tyson}, {Dell'Antonio},
  {Becker}, {Margoniner}, {Cohen}, {Norman}, {Loomba}, {Squires}, {Wilson},
  {Stubbs}, {Hennawi}, {Spergel}, {Boeshaar}, {Clocchiatti}, {Hamuy},
  {Bernstein}, {Gonzalez}, {Guhathakurta}, {Hu}, {Seljak}, \&
  {Zaritsky}}]{Witt:02}
{Wittman} D.~M. {et~al.}, 2002, in Society of Photo-Optical Instrumentation
  Engineers (SPIE) Conference Series, Vol. 4836, Society of Photo-Optical
  Instrumentation Engineers (SPIE) Conference Series, {Tyson} J.~A., {Wolff}
  S., eds., pp. 73--82

\bibitem[{{York} {et~al}\mbox{.}(2000){York}, {Adelman}, {Anderson},
  {Anderson}, {Annis}, {Bahcall}, {Bakken}, {Barkhouser}, {Bastian}, {Berman},
  {Boroski}, {Bracker}, {Briegel}, {Briggs}, {Brinkmann}, {Brunner}, {Burles},
  {Carey}, {Carr}, {Castander}, {Chen}, {Colestock}, {Connolly}, {Crocker},
  {Csabai}, {Czarapata}, {Davis}, {Doi}, {Dombeck}, {Eisenstein}, {Ellman},
  {Elms}, {Evans}, {Fan}, {Federwitz}, {Fiscelli}, {Friedman}, {Frieman},
  {Fukugita}, {Gillespie}, {Gunn}, {Gurbani}, {de Haas}, {Haldeman}, {Harris},
  {Hayes}, {Heckman}, {Hennessy}, {Hindsley}, {Holm}, {Holmgren}, {Huang},
  {Hull}, {Husby}, {Ichikawa}, {Ichikawa}, {Ivezi{\'c}}, {Kent}, {Kim},
  {Kinney}, {Klaene}, {Kleinman}, {Kleinman}, {Knapp}, {Korienek}, {Kron},
  {Kunszt}, {Lamb}, {Lee}, {Leger}, {Limmongkol}, {Lindenmeyer}, {Long},
  {Loomis}, {Loveday}, {Lucinio}, {Lupton}, {MacKinnon}, {Mannery}, {Mantsch},
  {Margon}, {McGehee}, {McKay}, {Meiksin}, {Merelli}, {Monet}, {Munn},
  {Narayanan}, {Nash}, {Neilsen}, {Neswold}, {Newberg}, {Nichol}, {Nicinski},
  {Nonino}, {Okada}, {Okamura}, {Ostriker}, {Owen}, {Pauls}, {Peoples},
  {Peterson}, {Petravick}, {Pier}, {Pope}, {Pordes}, {Prosapio},
  {Rechenmacher}, {Quinn}, {Richards}, {Richmond}, {Rivetta}, {Rockosi},
  {Ruthmansdorfer}, {Sandford}, {Schlegel}, {Schneider}, {Sekiguchi}, {Sergey},
  {Shimasaku}, {Siegmund}, {Smee}, {Smith}, {Snedden}, {Stone}, {Stoughton},
  {Strauss}, {Stubbs}, {SubbaRao}, {Szalay}, {Szapudi}, {Szokoly}, {Thakar},
  {Tremonti}, {Tucker}, {Uomoto}, {Vanden Berk}, {Vogeley}, {Waddell}, {Wang},
  {Watanabe}, {Weinberg}, {Yanny}, {Yasuda}, \& {SDSS Collaboration}}]{York:00}
{York} D.~G. {et~al.}, 2000, \aj, 120, 1579

\end{thebibliography}

\end{document}